\documentstyle[12pt]{article}
\newcommand{\definition}{\stackrel{\rm d}{\equiv}}
\begin{document}
\begin{titlepage}
\begin{flushright}
February 1995
\vspace{3cm}\\
\end{flushright}
\begin{center}
{\bf\Huge Gauging Higher Derivatives}
\vspace{4cm}\\
{\Large Shinji HAMAMOTO}
\vspace{1cm}\\
{\it Department of Physics, Toyama University \\
Toyama 930, JAPAN}
\vspace{3cm}\\
\end{center}
\begin{abstract}
The usual prescription for constructing gauge-invariant Lagrangian
is generalized to the case where a Lagrangian contains
second derivatives of fields as well as first derivatives.
Symmetric tensor fields in addition to the usual vector fields are
introduced as gauge fields.
Covariant derivatives and gauge-field strengths are determined.
\end{abstract}
\end{titlepage}

\section{Introduction}

Higher-derivative theories have been investigated in various contexts
of physics: higher-derivative terms naturally occur as quantum corrections
to lower-derivative theories;
nonlocal theories, for example string theories, are considered to be
equivalent to higher-derivative theories;
gravity theories with $R^{2}$ terms are good candidates to renormalizable
theories of gravity.${}^{1)}$

It is pointed out, however, that higher-derivative theories
have some crucial disadvantages:${}^{2),3),4)}$
they are unbounded from below;
they violate unitarity;
they break down the initial value problem.
Several attempts have been made to remove these faults,
but without success.

In the present paper we generalize the usual prescription for constructing
gauge-invariant Lagrangian to the case where higher-derivatives of fields
are included.
We expect that this new type of gauge degeneracy could be useful
to settle the above-mentioned problems of higher-derivative theories.

In \S 2 we consider a generic action that contains arbitrary order
derivatives of fields.
Requiring the action be invariant
under some not only global but also local transformations,
we get a series of identities.
These are generalizations of the usual Noether's theorems.
In \S 3 and the followings we restrict ourselves to treating
a simple case where the action has at most second derivatives.
Tensor gauge fields associated with second derivatives are introduced.
A series of identities that the tensor gauge fields as well as
the usual vector gauge fields should satisfy are writtten down in \S 3.
In \S 4 first and second derivatives of the two kinds of gauge fields
are decomposed to their irreducible components.
In \S 5 we  solve the identities to get covariant derivatives
and gauge-field strengths.
Summary and discussion are briefly given in \S 6.

\section{Identities}

In this section we consider a generic higher-derivative theories
whose action depends on fields $\varphi_{A}(x)$ and their derivatives
up to $N$th order
\begin{equation}
S
\stackrel{\rm d}{\equiv}\int {\rm d}^{4}x
{\cal L}(\varphi_{A}, \partial_{\mu}\varphi_{A},
\partial_{\mu\nu}\varphi_{A}, \cdots, \partial_{\mu_{1}\mu_{2}\cdots
\mu_{N}}\varphi_{A}) ,
\end{equation}
where we have used the abbreviation
\begin{equation}
\partial_{\mu_{1}\mu_{2}\cdots\mu_{n}}\varphi_{A}
\stackrel{\rm d}{\equiv}
\partial_{\mu_{1}}\partial_{\mu_{2}}\cdots\partial_{\mu_{n}}\varphi_{A} .
\end{equation}
The Euler-Lagrange equations are
\begin{equation}
\frac{\delta S}{\delta\varphi_{A}}
\stackrel{\rm d}{\equiv}
\sum_{n=0}^{N}(-)^{n}
\partial_{\mu_{1}\cdots\mu_{n}}
(\partial^{A,\mu_{1}\cdots\mu_{n}}{\cal L}) = 0 .
\end{equation}
Here and hereafter the differentiation
\begin{equation}
\partial^{A,\mu_{1}\cdots\mu_{n}}{\cal L}
\stackrel{\rm d}{\equiv}
\frac{\partial {\cal L}}{\partial\partial_{\mu_{1}\cdots\mu_{n}}\varphi_{A}}
\end{equation}
always stands for that of weight 1:
in the case of ${\cal L} = C^{A\mu\nu}\partial_{\mu\nu}\varphi_{A}$,
for example, we have
$\partial^{A,\mu\nu}{\cal L} = \frac{1}{2!}( C^{A\mu\nu} + C^{A\nu\mu} )$.

Suppose the action is invariant under a ^^ global' infinitesimal
transformation with $p$ independent parameters $\epsilon^{a}$
\begin{equation}
\left\{
\begin{array}{rcl}
\delta x^{\mu}       & = & \epsilon^{a}X^{\mu}_{a}(x) , \\
\delta\varphi_{A}(x) & = & \epsilon^{a}N_{aA}(x,\varphi) .
\end{array}
\right.
\end{equation}
Then $p$ identities follow:${}^{5)}$
\begin{eqnarray}
\lefteqn{
\frac{\delta S}{\delta\varphi_{A}}
\left( N_{aA} - \partial_{\nu}\varphi_{A}\cdot X^{\nu}_{a} \right)
}\nonumber \\
& + & \partial_{\mu}\left[ \sum_{n=0}^{N-1}\left(
\partial^{A, \mu\alpha_{1}\cdots\alpha_{n}}{\cal L} \right)
\stackrel{\leftrightarrow}{\partial}_{\alpha_{1}\cdots\alpha_{n}}
\left( N_{aA} - \partial_{\nu}\varphi_{A}\cdot X^{\nu}_{a} \right)
+ {\cal L}X^{\mu}_{a} \right] \equiv 0 ,
\end{eqnarray}
where the generalized ^^ both-side' derivatives for two arbitrary
functions $F$ and $G$
\begin{equation}
F\stackrel{\leftrightarrow}{\partial}_{\mu_{1}\cdots\mu_{n}}G
\stackrel{\rm d}{\equiv}
\sum_{k=0}^{n}(-)^{k}\partial_{\mu_{1}\cdots\mu_{k}}F\cdot
\partial_{\mu_{k+1}\cdots\mu_{n}}G
\end{equation}
has been introduced.
These identities are also written in the form of
\begin{equation}
\sum_{n=0}^{N}\partial^{A,\alpha_{1}\cdots\alpha_{n}}{\cal L}\cdot
\partial_{\alpha_{1}\cdots\alpha_{n}}
\left( N_{aA} - \partial_{\mu}\varphi_{A}\cdot X^{\mu}_{a} \right)
+ \partial_{\mu}\left( {\cal L}X^{\mu}_{a} \right) \equiv 0 .
\end{equation}
Generalized Noether currents are given by the bracketed quantities
in Eq.(6):
\begin{equation}
J^{\mu}_{a} \stackrel{\rm d}{\equiv}
\sum_{n=0}^{N-1}\left(
\partial^{A, \mu\alpha_{1}\cdots\alpha_{n}}{\cal L} \right)
\stackrel{\leftrightarrow}{\partial}_{\alpha_{1}\cdots\alpha_{n}}
\left( N_{aA} - \partial_{\nu}\varphi_{A}\cdot X^{\nu}_{a} \right)
+ {\cal L}X^{\mu}_{a} ,
\end{equation}
which are conserved when the Euler-Lagrange equations
are satisfied
\begin{equation}
\partial_{\mu}J^{\mu}_{a} = 0 .
\end{equation}

Let us now proceed to ^^ local' infinitesimal transformation.
In the present case,
since the action contains up-to-$N$th order derivatives,
the local version of the transformation (5) can be given as follows:
\begin{equation}
\left\{
\begin{array}{rcl}
\delta x^{\mu}       & = & \lambda^{a}(x)X^{\mu}_{a}(x) , \\
\delta\varphi_{A}(x) & =
 & \displaystyle\sum_{n=0}^{N} \partial_{\mu_{1}\cdots\mu_{n}}\lambda^{a}(x)
    \cdot N^{\mu_{1}\cdots\mu_{n}}_{aA}(x,\varphi) ,
\end{array}
\right.
\end{equation}
where $\lambda^{a}(x)$ are $p$ independent arbitrary functions.
If we require the action be invariant under the local (infinitesimal)
transformation (11), we get the following series of identities:
\begin{eqnarray}
\lefteqn{
\sum_{n=0}^{N}\partial^{A,\alpha_{1}\cdots\alpha_{n}}{\cal L}\cdot
\partial_{\alpha_{1}\cdots\alpha_{n}}
\left( N_{aA} - \partial_{\mu}\varphi_{A}\cdot X^{\mu}_{a} \right)
+ \partial_{\mu}\left( {\cal L}X^{\mu}_{a} \right) \equiv 0 , }\\
\lefteqn{
\sum_{n=1}^{N}{}_{n}C_{1}
\partial^{A, \nu_{1}\alpha_{2}\cdots\alpha_{n}}{\cal L}
\cdot\partial_{\alpha_{2}\cdots\alpha_{n}}
\left( N_{aA} - \partial_{\mu}\varphi_{A}\cdot X^{\mu}_{a} \right)
}\nonumber \\
& + & \sum_{n=0}^{N} \partial^{A, \alpha_{1}\cdots\alpha_{n}}{\cal L}\cdot
\partial_{\alpha_{1}\cdots\alpha_{n}}N^{\nu_{1}}_{aA}
+ {\cal L}X^{\nu_{1}}_{a} \equiv 0 , \\
\lefteqn{
\sum_{n=2}^{N}{}_{n}C_{2}
\partial^{A,\nu_{1}\nu_{2}\alpha_{3}\cdots\alpha_{n}}{\cal L}
\cdot\partial_{\alpha_{3}\cdots\alpha_{n}}
\left( N_{aA} - \partial_{\mu}\varphi_{A}\cdot X^{\mu}_{a} \right)
}\nonumber \\
& + & \frac{1}{2}\sum_{n=1}^{N}{}_{n}C_{1}
\left[ \partial^{A,\nu_{1}\alpha_{2}\cdots\alpha_{n}}{\cal L}\cdot
\partial_{\alpha_{2}\cdots\alpha_{n}}N^{\nu_{2}}_{aA}
+ (\nu_{1} \leftrightarrow \nu_{2} ) \right]
\nonumber \\
& + & \sum_{n=0}^{N} \partial^{A,\alpha_{1}\cdots\alpha_{n}}{\cal L}\cdot
\partial_{\alpha_{1}\cdots\alpha_{n}}N^{\nu_{1}\nu_{2}}_{aA}
\equiv 0 , \\
\lefteqn{
\sum_{n=3}^{N}{}_{n}C_{3}
\partial^{A,\nu_{1}\nu_{2}\nu_{3}\alpha_{4}\cdots\alpha_{n}}{\cal L}\cdot
\partial_{\alpha_{4}\cdots\alpha_{n}}
\left( N_{aA} - \partial_{\mu}\varphi_{A}\cdot X^{\mu}_{a}\right)
}\nonumber \\
& + & \frac{1}{3}\sum_{n=2}^{N}{}_{n}C_{2}
\left[ \partial^{A,\nu_{1}\nu_{2}\alpha_{3}\cdots\alpha_{n}}{\cal L}\cdot
\partial_{\alpha_{3}\cdots\alpha_{n}}N^{\nu_{3}}_{aA}
+ ( \nu_{1}\nu_{2}\nu_{3}\mbox{--cyclic} )\right]
\nonumber \\
& + & \frac{1}{3}\sum_{n=1}^{N}{}_{n}C_{1}
\left[ \partial^{A,\nu_{1}\alpha_{2}\cdots\alpha_{n}}{\cal L}\cdot
\partial_{\alpha_{2}\cdots\alpha_{n}}N^{\nu_{2}\nu_{3}}_{aA}
+ ( \nu_{1}\nu_{2}\nu_{3}\mbox{--cyclic} )\right]
\nonumber \\
& + & \sum_{n=0}^{N}\partial^{A,\alpha_{1}\cdots\alpha_{n}}{\cal L}\cdot
\partial_{\alpha_{1}\cdots\alpha_{n}}N^{\nu_{1}\nu_{2}\nu_{3}}_{aA}
\equiv 0 , \\
\lefteqn{} \nonumber \\
\lefteqn{\vdots} \nonumber \\
\lefteqn{} \nonumber \\
\lefteqn{
\partial^{A,\nu_{1}\cdots\nu_{N}}{\cal L}\cdot
\left( N_{aA} - \partial_{\mu}\varphi_{A}\cdot X^{\mu}_{a}\right)
}\nonumber \\
& + & \frac{1}{{}_{N}C_{N-1}}\sum_{n=N-1}^{N}{}_{n}C_{N-1}
\nonumber \\
& & \makebox[15mm]{}
\times \left[ \partial^{A,\nu_{1}\cdots\nu_{N-1}\alpha_{N}\alpha_{n}}
{\cal L}\cdot
\partial_{\alpha_{N}\alpha_{n}}N^{\nu_{N}}_{aA}
+ ( \nu_{1}\cdots\nu_{N}\mbox{--combination} )\right]
\nonumber \\
& + & \cdots \nonumber \\
& + & \frac{1}{{}_{N}C_{1}}\sum_{n=1}^{N}{}_{n}C_{1}
\left[ \partial^{A,\nu_{1}\alpha_{2}\cdots\alpha_{n}}{\cal L}\cdot
\partial_{\alpha_{2}\cdots\alpha_{n}}N^{\nu_{2}\cdots\nu_{N}}_{aA}
+ ( \nu_{1}\cdots\nu_{N}\mbox{--combination} )\right]
\nonumber \\
& + & \sum_{n=0}^{N} \partial^{A,\alpha_{1}\cdots\alpha_{n}}{\cal L}\cdot
\partial_{\alpha_{1}\cdots\alpha_{n}}N^{\nu_{1}\cdots\nu_{N}}_{aA}
\equiv 0 , \\
\lefteqn{
\frac{1}{{}_{N+1}C_{N}}
\left[ \partial^{A,\nu_{1}\cdots\nu_{N}}{\cal L}\cdot N^{\nu_{N+1}}_{aA}
+ ( \nu_{1}\cdots\nu_{N+1}\mbox{--combination} )\right]
}\nonumber \\
& + & \frac{1}{{}_{N+1}C_{N-1}}\sum_{n=N-1}^{N}{}_{n}C_{N-1}
\nonumber \\
& & \makebox[15mm]{}
\times \left[ \partial^{A,\nu_{1}\cdots\nu_{N-1}\alpha_{N}\alpha_{n}}
{\cal L}\cdot
\partial_{\alpha_{N}\alpha_{n}}N^{\nu_{N}\nu_{N+1}}_{aA}
+ ( \nu_{1}\cdots\nu_{N+1}\mbox{--combination} )\right]
\nonumber \\
& + & \cdots \nonumber \\
& + & \frac{1}{{}_{N+1}C_{1}}\sum_{n=1}^{N}{}_{n}C_{1}
\left[ \partial^{A,\nu_{1}\alpha_{2}\cdots\alpha_{n}}{\cal L}\cdot
\partial_{\alpha_{2}\cdots\alpha_{n}}N^{\nu_{2}\cdots\nu_{N+1}}_{aA}
+ ( \nu_{1}\cdots\nu_{N+1}\mbox{--combination} )\right]
\nonumber \\
& \equiv & 0 , \\
\lefteqn{} \nonumber \\
\lefteqn{\vdots} \nonumber \\
\lefteqn{} \nonumber \\
\lefteqn{
\frac{1}{{}_{2N-1}C_{N}}
\left[ \partial^{A,\nu_{1}\cdots\nu_{N}}{\cal L}\cdot
N^{\nu_{N+1}\cdots\nu_{2N-1}}_{aA}
+ ( \nu_{1}\cdots\nu_{2N-1}\mbox{--combination} )\right]
}\nonumber \\
& + & \frac{1}{{}_{2N-1}C_{N-1}}\sum_{n=N-1}^{N}{}_{n}C_{N-1}
\nonumber \\
& & \makebox[15mm]{}
\times \left[ \partial^{A,\nu_{1}\cdots\nu_{N-1}\alpha_{N}\alpha_{n}}
{\cal L}\cdot
\partial_{\alpha_{N}\alpha_{n}}N^{\nu_{N}\cdots\nu_{2N-1}}_{aA}
+ ( \nu_{1}\cdots\nu_{2N-1}\mbox{--combination} )\right]
\nonumber \\
& \equiv & 0 , \\
\lefteqn{
\frac{1}{{}_{2N}C_{N}}
\left[ \partial^{A,\nu_{1}\cdots\nu_{N}}{\cal L}\cdot
N^{\nu_{N+1}\cdots\nu_{2N}}_{aA}
+ ( \nu_{1}\cdots\nu_{2N}\mbox{--combination} )\right]
\equiv 0 .}
\end{eqnarray}
The identities (12) are the same as (8),
obtained as the coefficients of $\lambda^{a}(x)$ in the requirement
$\delta S \equiv 0$.
The coefficients of $\partial_{\nu_{1}}\lambda^{a},
                     \partial_{\nu_{1}\nu_{2}}\lambda^{a},
                     \cdots$, and
                    $\partial_{\nu_{1}\cdots\nu_{2N}}\lambda^{a}$
give (13), (14), $\cdots$, and (19) respectively.

\section{Gauge fields}

In this and the following sections, we set $N = 2$
for the sake of simplicity.
The starting action is therefore
\begin{equation}
S \stackrel{\rm d}{\equiv}
\int {\rm d}^{4}x {\cal L} \left( \varphi_{A}, \partial_{\mu}\varphi_{A},
\partial_{\mu\nu}\varphi_{A} \right) .
\end{equation}
This action is assumed to be invariant
under the global {\it internal} transformation
\begin{equation}
\left\{
\begin{array}{rcl}
\delta x^{\mu}       & = & 0 , \\
\delta\varphi_{A}(x) & =
                  & \epsilon^{a}M_{aA}^{\makebox[3mm]{} B}\varphi_{B}(x) ,
\end{array}
\right.
\end{equation}
where $M_{aA}^{\makebox[3mm]{} B}$
are certain representation matrices of a group $G$.
That means the Lagrangian ${\cal L}$ should satisfy the following identities:
\begin{equation}
\partial^{A}{\cal L}\cdot M_{aA}^{\makebox[3mm]{} B}\varphi_{B}
+ \partial^{A,\mu}{\cal L}\cdot
M_{aA}^{\makebox[3mm]{} B}\partial_{\mu}\varphi_{B}
+ \partial^{A,\mu\nu}{\cal L}\cdot
M_{aA}^{\makebox[3mm]{} B}\partial_{\mu\nu}\varphi_{B}
\equiv 0 .
\end{equation}

Next consider the local version of the transformation (21).
The transformation obtained by simply replacing
the arbitrary parameters $\epsilon^{a}$
with arbitrary function $\lambda^{a}(x)$
\begin{equation}
\left\{
\begin{array}{rcl}
\delta x^{\mu}       & = & 0 , \\
\delta\varphi_{A}(x) & =
           & \lambda^{a}(x)M_{aA}^{\makebox[3mm]{} B}\varphi_{B}(x)
\end{array}
\right.
\end{equation}
does not leave the action invariant.
Generalizing the usual prescription for gaugeization,
we introduce two kinds of gauge fields $B^{a}_{\mu}(x)$ and
$B^{a}_{\mu\nu}(x)$:
the vector fields $B^{a}_{\mu}(x)$ are to be combined with the first
derivative $\partial_{\mu}$ as usual;
the symmetric tensor fields $B^{a}_{\mu\nu}(x)$
are newly introduced to form second-order covariant derivative.
It seems natural to assume the following transformation properties for
these gauge fields:
\begin{equation}
\left\{
\begin{array}{rcl}
\delta B^{a}_{\mu}(x) & =
& \lambda^{b}(x)f^{a}_{bc}B^{c}_{\mu}(x) + \partial_{\mu}\lambda^{a}(x) , \\
\delta B^{a}_{\mu\nu}(x) & =
& \displaystyle \lambda^{b}(x)f^{a}_{bc}B^{c}_{\mu\nu}(x)
+ \frac{1}{2} \left[
\partial_{\mu}\lambda^{b}(x)f^{a}_{bc}B^{c}_{\nu}(x)
+ ( \mu \leftrightarrow \nu ) \right]
+ \partial_{\mu\nu}\lambda^{a}(x) , \\
\end{array}
\right.
\end{equation}
where $f^{a}_{bc}$ are structure constants of $G$
\begin{equation}
\left[ M_{b} , M_{c} \right] = f^{a}_{bc}M_{a} .
\end{equation}
The action $S$ should be extended to incorporate the gauge fields
{\it and} their derivatives up to second order:
\begin{equation}
S_{1} \stackrel{\rm d}{\equiv}
\int {\rm d}^{4}x {\cal L}_{1}
\left( \varphi_{A}, \partial_{\mu}\varphi_{A}, \partial_{\mu\nu}\varphi_{A};
B^{a}_{\mu}, \partial_{\mu}B^{a}_{\nu}, \partial_{\mu\nu}B^{a}_{\rho};
B^{a}_{\mu\nu}, \partial_{\mu}B^{a}_{\nu\rho},
\partial_{\mu\nu}B^{a}_{\rho\sigma} \right).
\end{equation}
In what form is the Lagrangian ${\cal L}_{1}$ to contain the gauge fields
and their derivatives?
To answer this question we require that
the action $S_{1}$ be invariant
under the local transformation (23) and (24).

The invariance requirement gives the following series of identities:
\begin{eqnarray}
\lefteqn{
\frac{\partial {\cal L}_{1}}{\partial\varphi_{A}}
M_{aA}^{\makebox[3mm]{} B}\varphi_{B}
+ \frac{\partial {\cal L}_{1}}{\partial\partial_{\mu}\varphi_{A}}
M_{aA}^{\makebox[3mm]{} B}\partial_{\mu}\varphi_{B}
+ \frac{\partial {\cal L}_{1}}{\partial\partial_{\mu\nu}\varphi_{A}}
M_{aA}^{\makebox[3mm]{} B}\partial_{\mu\nu}\varphi_{B}
}\nonumber \\
& + & \frac{\partial {\cal L}_{1}}{\partial B^{b}_{\rho}}
f^{b}_{ac}B^{c}_{\rho}
+ \frac{\partial {\cal L}_{1}}{\partial\partial_{\mu}B^{b}_{\rho}}
f^{b}_{ac}\partial_{\mu}B^{c}_{\rho}
+ \frac{\partial {\cal L}_{1}}{\partial\partial_{\mu\nu}B^{b}_{\rho}}
f^{b}_{ac}\partial_{\mu\nu}B^{c}_{\rho}
\nonumber \\
& + & \frac{\partial {\cal L}_{1}}{\partial B^{b}_{\rho\sigma}}
f^{b}_{ac}B^{c}_{\rho\sigma}
+ \frac{\partial {\cal L}_{1}}{\partial\partial_{\mu}B^{b}_{\rho\sigma}}
f^{b}_{ac}\partial_{\mu}B^{c}_{\rho\sigma}
+ \frac{\partial {\cal L}_{1}}{\partial\partial_{\mu\nu}B^{b}_{\rho\sigma}}
f^{b}_{ac}\partial_{\mu\nu}B^{c}_{\rho\sigma}
\equiv 0 , \\
\lefteqn{
\frac{\partial {\cal L}_{1}}{\partial\partial_{\alpha}\varphi_{A}}
M_{aA}^{\makebox[3mm]{} B}\varphi_{B}
+ 2 \frac{\partial {\cal L}_{1}}{\partial\partial_{\alpha\mu}\varphi_{A}}
M_{aA}^{\makebox[3mm]{} B}\partial_{\mu}\varphi_{B}
}\nonumber \\
& + & \frac{\partial {\cal L}_{1}}{\partial\partial_{\alpha}B^{b}_{\rho}}
f^{b}_{ac}B^{c}_{\rho}
+ 2 \frac{\partial {\cal L}_{1}}{\partial\partial_{\alpha\mu}B^{b}_{\rho}}
f^{b}_{ac}\partial_{\mu}B^{c}_{\rho}
\nonumber \\
& + & \frac{\partial {\cal L}_{1}}
{\partial\partial_{\alpha}B^{b}_{\rho\sigma}}
f^{b}_{ac}B^{c}_{\rho\sigma}
+ 2 \frac{\partial {\cal L}_{1}}
{\partial\partial_{\alpha\mu}B^{b}_{\rho\sigma}}
f^{b}_{ac}\partial_{\mu}B^{c}_{\rho\sigma}
+ \frac{\partial {\cal L}_{1}}{\partial B^{a}_{\alpha}}
\nonumber \\
& + & \frac{\partial {\cal L}_{1}}{\partial B^{b}_{\alpha\rho}}
f^{b}_{ac}B^{c}_{\rho}
+ \frac{\partial {\cal L}_{1}}{\partial\partial_{\mu}B^{b}_{\alpha\rho}}
f^{b}_{ac}\partial_{\mu}B^{c}_{\rho}
+ \frac{\partial {\cal L}_{1}}{\partial\partial_{\mu\nu}B^{b}_{\alpha\rho}}
f^{b}_{ac}\partial_{\mu\nu}B^{c}_{\rho}
\equiv 0 , \\
\lefteqn{
\frac{\partial {\cal L}_{1}}{\partial\partial_{\alpha\beta}\varphi_{A}}
M_{aA}^{\makebox[3mm]{} B}\varphi_{B}
}\nonumber \\
& + & \frac{\partial {\cal L}_{1}}
{\partial\partial_{\alpha\beta}B^{b}_{\rho}}
f^{b}_{ac}B^{c}_{\rho}
+ \frac{\partial {\cal L}_{1}}
{\partial\partial_{\alpha\beta}B^{b}_{\rho\sigma}}
f^{b}_{ac}B^{c}_{\rho\sigma}
+ \frac{1}{2} \left(
\frac{\partial {\cal L}_{1}}{\partial\partial_{\alpha}B^{a}_{\beta}}
+ \frac{\partial {\cal L}_{1}}
{\partial\partial_{\beta}B^{a}_{\alpha}} \right)
+ \frac{\partial {\cal L}_{1}}{\partial B^{a}_{\alpha\beta}}
\nonumber \\
& + & \frac{1}{2} \left(
\frac{\partial {\cal L}_{1}}{\partial\partial_{\alpha}B^{b}_{\beta\rho}}
+ \frac{\partial {\cal L}_{1}}
{\partial\partial_{\beta}B^{b}_{\alpha\rho}} \right)
f^{b}_{ac}B^{c}_{\rho}
+ \left(
\frac{\partial {\cal L}_{1}}{\partial\partial_{\alpha\mu}B^{b}_{\beta\rho}}
+ \frac{\partial {\cal L}_{1}}{\partial\partial_{\beta\mu}B^{b}_{\alpha\rho}}
\right)
f^{b}_{ac}\partial_{\mu}B^{c}_{\rho}
\equiv 0 , \\
\lefteqn{
\frac{1}{3} \left(
\frac{\partial {\cal L}_{1}}{\partial\partial_{\alpha\beta}B^{a}_{\gamma}}
+ \frac{\partial {\cal L}_{1}}{\partial\partial_{\beta\gamma}B^{a}_{\alpha}}
+ \frac{\partial {\cal L}_{1}}
{\partial\partial_{\gamma\alpha}B^{a}_{\beta}} \right)
+ \frac{1}{3} \left(
\frac{\partial {\cal L}_{1}}{\partial\partial_{\alpha}B^{a}_{\beta\gamma}}
+ \frac{\partial {\cal L}_{1}}{\partial\partial_{\beta}B^{a}_{\gamma\alpha}}
+ \frac{\partial {\cal L}_{1}}
{\partial\partial_{\gamma}B^{a}_{\alpha\beta}} \right)
}\nonumber \\
& + & \frac{1}{3}\left(
\frac{\partial {\cal L}_{1}}
{\partial\partial_{\alpha\beta}B^{b}_{\gamma\rho}}
+ \frac{\partial {\cal L}_{1}}
{\partial\partial_{\beta\gamma}B^{b}_{\alpha\rho}}
+ \frac{\partial {\cal L}_{1}}
{\partial\partial_{\gamma\alpha}B^{b}_{\beta\rho}}
\right)
f^{b}_{ac}B^{c}_{\rho}
\equiv 0 , \\
\lefteqn{
\frac{1}{6} \left(
\frac{\partial {\cal L}_{1}}
{\partial\partial_{\alpha\beta}B^{a}_{\gamma\delta}}
+ \frac{\partial {\cal L}_{1}}
{\partial\partial_{\alpha\gamma}B^{a}_{\beta\delta}}
+ \frac{\partial {\cal L}_{1}}
{\partial\partial_{\alpha\delta}B^{a}_{\beta\gamma}}
+ \frac{\partial {\cal L}_{1}}
{\partial\partial_{\gamma\delta}B^{a}_{\alpha\beta}}
+ \frac{\partial {\cal L}_{1}}
{\partial\partial_{\beta\delta}B^{a}_{\alpha\gamma}}
+ \frac{\partial {\cal L}_{1}}
{\partial\partial_{\beta\gamma}B^{a}_{\alpha\delta}}
\right)
}\nonumber \\
& \equiv & 0 .
\end{eqnarray}
Solving these identities will determine the forms of covariant derivatives
and gauge-field strengths,
through which the gauge fields and their derivatives are contained in the
Lagrangian ${\cal L}_{1}$.
This is the task of the next two sections.

\section{Irreducible decomposition}

To solve the identities it is useful to decompose the derivatives of
the gauge fields into their irreducible components.
This is done by using Young's prescription
and by taking into account symmetric properties such as
$\partial_{\mu\nu} = \partial_{\nu\mu}$ and
$B^{a}_{\mu\nu} = B^{a}_{\nu\mu}$.

The first and second derivatives of $B^{a}_{\mu}$ are decomposed as
\begin{eqnarray}
F^{(i)}_{\alpha\beta} & \definition
 & \Phi ^{(i)\mu\nu}_{\alpha\beta}\partial_{\mu}B_{\nu} ,
   \makebox[1cm]{} ( i = 1, 2 ) \\
C^{(i)}_{\alpha\beta\gamma} & \definition
 & \Gamma ^{(i)\lambda\mu\nu}_{\alpha\beta\gamma}
   \partial_{\lambda\mu}B_{\nu} ,
   \makebox[1cm]{} ( i = 1, 2 )
\end{eqnarray}
where
\begin{equation}
\left\{
\begin{array}{rclll}
\Phi ^{(1)\mu\nu}_{\alpha\beta} & \definition & \displaystyle
 \frac{1}{2}\left( \delta^{\mu}_{\alpha}\delta^{\nu}_{\beta}
                 + \delta^{\mu}_{\beta}\delta^{\nu}_{\alpha} \right)
 & \definition & \delta^{\mu\nu}_{\alpha\beta} , \\
\Phi ^{(2)\mu\nu}_{\alpha\beta} & \definition & \displaystyle
 \frac{1}{2}\left( \delta^{\mu}_{\alpha}\delta^{\nu}_{\beta}
                 - \delta^{\mu}_{\beta}\delta^{\nu}_{\alpha} \right) ;
 & &
\end{array}
\right.
\end{equation}
and
\begin{equation}
\left\{
\begin{array}{rcl}
\Gamma ^{(1)\lambda\mu\nu}_{\alpha\beta\gamma} & \definition & \displaystyle
 \frac{1}{3}\left( \delta^{\lambda\mu}_{\alpha\beta}\delta^{\nu}_{\gamma}
                 + \delta^{\lambda\mu}_{\beta\gamma}\delta^{\nu}_{\alpha}
                 + \delta^{\lambda\mu}_{\gamma\alpha}\delta^{\nu}_{\beta}
 \right) , \\
\Gamma ^{(2)\lambda\mu\nu}_{\alpha\beta\gamma} & \definition & \displaystyle
 \frac{1}{4}\left( 2\delta^{\lambda\mu}_{\alpha\beta}\delta^{\nu}_{\gamma}
                  - \delta^{\lambda\mu}_{\beta\gamma}\delta^{\nu}_{\alpha}
                  - \delta^{\lambda\mu}_{\gamma\alpha}\delta^{\nu}_{\beta}
 \right) .
\end{array}
\right.
\end{equation}
The first and second derivatives of $B^{a}_{\mu\nu}$ are decomposed as
\begin{eqnarray}
D^{(i)}_{\alpha\beta\gamma} & \definition
 & \Delta ^{(i)\lambda\mu\nu}_{\alpha\beta\gamma}
   \partial_{\lambda}B_{\mu\nu} , \makebox[1cm]{} ( i = 1, 2 ) \\
G^{(i)}_{\alpha\beta\gamma\delta} & \definition
 & \Xi ^{(i)\kappa\lambda\mu\nu}_{\alpha\beta\gamma\delta}
   \partial_{\kappa\lambda}B_{\mu\nu} , \makebox[1cm]{} ( i = 1, 2, 3 )
\end{eqnarray}
where
\begin{equation}
\left\{
\begin{array}{rcl}
\Delta ^{(1)\lambda\mu\nu}_{\alpha\beta\gamma} & \definition
 & \Gamma ^{(1)\mu\nu\lambda}_{\alpha\beta\gamma} , \\
\Delta ^{(2)\lambda\mu\nu}_{\alpha\beta\gamma} & \definition
 & \Gamma ^{(2)\mu\nu\lambda}_{\alpha\beta\gamma} ;
\end{array}
\right.
\end{equation}
and
\begin{equation}
\left\{
\begin{array}{rcl}
\Xi ^{(1)\kappa\lambda\mu\nu}_{\alpha\beta\gamma\delta}
 & \definition
 & \displaystyle \frac{1}{6}
   \left[ \left( \delta^{\kappa\lambda}_{\alpha\beta}
                 \delta^{\mu\nu}_{\gamma\delta}
               + \delta^{\kappa\lambda}_{\gamma\delta}
                 \delta^{\mu\nu}_{\alpha\beta} \right)
        + \left( \delta^{\kappa\lambda}_{\alpha\gamma}
                 \delta^{\mu\nu}_{\beta\delta}
               + \delta^{\kappa\lambda}_{\beta\delta}
                 \delta^{\mu\nu}_{\alpha\gamma} \right)
        + \left( \delta^{\kappa\lambda}_{\beta\gamma}
                 \delta^{\mu\nu}_{\alpha\delta}
               + \delta^{\kappa\lambda}_{\alpha\delta}
                 \delta^{\mu\nu}_{\beta\gamma} \right) \right] , \\
\Xi ^{(2)\kappa\lambda\mu\nu}_{\alpha\beta\gamma\delta}
 & \definition
 & \displaystyle \frac{1}{6}
   \left[ \left( \delta^{\kappa\lambda}_{\alpha\beta}
                 \delta^{\mu\nu}_{\gamma\delta}
               - \delta^{\kappa\lambda}_{\gamma\delta}
                 \delta^{\mu\nu}_{\alpha\beta} \right)
        + \left( \delta^{\kappa\lambda}_{\alpha\gamma}
                 \delta^{\mu\nu}_{\beta\delta}
               - \delta^{\kappa\lambda}_{\beta\delta}
                 \delta^{\mu\nu}_{\alpha\gamma} \right)
        + \left( \delta^{\kappa\lambda}_{\beta\gamma}
                 \delta^{\mu\nu}_{\alpha\delta}
               - \delta^{\kappa\lambda}_{\alpha\delta}
                 \delta^{\mu\nu}_{\beta\gamma} \right) \right] , \\
\Xi ^{(3)\kappa\lambda\mu\nu}_{\alpha\beta\gamma\delta}
 & \definition
 & \displaystyle \frac{1}{8}
   \left[ 2\left( \delta^{\kappa\lambda}_{\alpha\beta}
                 \delta^{\mu\nu}_{\gamma\delta}
               + \delta^{\kappa\lambda}_{\gamma\delta}
                 \delta^{\mu\nu}_{\alpha\beta} \right)
        - \left( \delta^{\kappa\lambda}_{\alpha\gamma}
                 \delta^{\mu\nu}_{\beta\delta}
               + \delta^{\kappa\lambda}_{\beta\delta}
                 \delta^{\mu\nu}_{\alpha\gamma} \right)
        - \left( \delta^{\kappa\lambda}_{\beta\gamma}
                 \delta^{\mu\nu}_{\alpha\delta}
               + \delta^{\kappa\lambda}_{\alpha\delta}
                 \delta^{\mu\nu}_{\beta\gamma} \right) \right] .
\end{array}
\right.
\end{equation}
The Lagrangian ${\cal L}_{1}$ is a function of these irreducible components:
\begin{equation}
{\cal L}_{1} \equiv
{\cal L}_{2}\left( \varphi, \partial_{\mu}\varphi, \partial_{\mu\nu}\varphi;
            B_{\rho}, F^{(1)}, F^{(2)}, C^{(1)}, C^{(2)};
            B_{\rho\sigma}, D^{(1)}, D^{(2)}, G^{(1)}, G^{(2)}, G^{(3)}
     \right) .
\end{equation}

\section{Covariant derivatives and field strengths}

For the Lagrangian ${\cal L}_{2}$ the identities (31) reduce to
\begin{equation}
\frac{\partial {\cal L}_{2}}{\partial G^{(1)a}_{\alpha\beta\gamma\delta}}
\equiv 0 .
\end{equation}
These identities tell the fact that the Lagrangian ${\cal L}_{2}$ is
independent of $G^{(1)}$:
\begin{equation}
{\cal L}_{2} \equiv
{\cal L}_{3}\left( \varphi, \partial_{\mu}\varphi, \partial_{\mu\nu}\varphi;
            B_{\rho}, F^{(1)}, F^{(2)}, C^{(1)}, C^{(2)};
            B_{\rho\sigma}, D^{(1)}, D^{(2)}, G^{(2)}, G^{(3)}
     \right) .
\end{equation}

The identities (30) are rewritten for the Lagrangian ${\cal L}_{3}$ as
\begin{eqnarray}
\lefteqn{
\frac{\partial {\cal L}_{3}}{\partial C^{(1)a}_{\alpha\beta\gamma}} +
\frac{\partial {\cal L}_{3}}{\partial D^{(1)a}_{\alpha\beta\gamma}} }
\nonumber \\
& & +
\frac{1}{3}\frac{\partial {\cal L}_{3}}{\partial G^{(2)b}_{\mu\nu\rho\sigma}}
f^{b}_{ac}\left( \Xi^{(2)\alpha\beta\gamma\delta}_{\mu\nu\rho\sigma} +
                 \Xi^{(2)\beta\gamma\alpha\delta}_{\mu\nu\rho\sigma} +
                 \Xi^{(2)\gamma\alpha\beta\delta}_{\mu\nu\rho\sigma} \right)
B^{c}_{\delta}
\equiv 0 .
\end{eqnarray}
These identities show that the Lagrangian ${\cal L}_{3}$ should
have $D^{(1)}$ in the form of
\begin{equation}
{\cal L}_{3} \equiv
{\cal L}_{4}\left( \varphi, \partial_{\mu}\varphi, \partial_{\mu\nu}\varphi;
            B_{\rho}, F^{(1)}, F^{(2)}, \tilde{C}^{(1)}, C^{(2)};
            B_{\rho\sigma}, D^{(2)}, \tilde{G}^{(2)}, G^{(3)}
     \right) ,
\end{equation}
where
\begin{eqnarray}
\tilde{C}^{(1)a}_{\alpha\beta\gamma}
& \definition &
C^{(1)a}_{\alpha\beta\gamma}
- D^{(1)a}_{\alpha\beta\gamma} , \\
\tilde{G}^{(2)a}_{\alpha\beta\gamma\delta}
& \definition &
G^{(2)a}_{\alpha\beta\gamma\delta}
- f^{a}_{bc}\Xi^{(2)\mu\nu\rho\sigma}_{\alpha\beta\gamma\delta}
  D^{(1)b}_{\mu\nu\rho}B^{c}_{\sigma} .
\end{eqnarray}

The Lagrangian ${\cal L}_{4}$ changes the form of the identities (29) into
\begin{eqnarray}
\lefteqn{
\frac{\partial {\cal L}_{4}}{\partial\partial_{\alpha\beta}\varphi_{A}}
M_{aA}^{\makebox[3mm]{} B}\varphi_{B}
}\nonumber \\
& + & \frac{\partial {\cal L}_{4}}{\partial C^{(2)b}_{\mu\nu\rho}}
      f^{b}_{ac}\Gamma^{(2)\alpha\beta\gamma}_{\mu\nu\rho}B^{c}_{\gamma}
    - \frac{1}{2}\frac{\partial {\cal L}_{4}}{\partial D^{(2)b}_{\mu\nu\rho}}
      f^{b}_{ac}\Delta^{(2)\gamma\alpha\beta}_{\mu\nu\rho}B^{c}_{\gamma}
\nonumber \\
& + & \frac{\partial {\cal L}_{4}}
{\partial \tilde{G}^{(2)b}_{\mu\nu\rho\sigma}}
      \left\{ f^{b}_{ac} \left[
      \Xi^{(2)\alpha\beta\gamma\delta}_{\mu\nu\rho\sigma}B^{c}_{\gamma\delta}
    + \left( \Xi^{(2)\alpha\gamma\beta\delta}_{\mu\nu\rho\sigma}
           + \Xi^{(2)\beta\gamma\alpha\delta}_{\mu\nu\rho\sigma} \right)
      F^{(2)c}_{\gamma\delta} \right] \right.
\nonumber \\
& & \makebox[45mm]{} \left. + f^{d}_{ea}f^{b}_{dc}
      \Xi^{(2)\xi\eta\zeta\upsilon}_{\mu\nu\rho\sigma}
      \Delta^{(1)\alpha\beta\gamma}_{\xi\eta\zeta}
      B^{c}_{\upsilon}B^{e}_{\gamma} \right\}
\nonumber \\
& + & \frac{\partial {\cal L}_{4}}{\partial G^{(3)b}_{\mu\nu\rho\sigma}}
      f^{b}_{ac}\Xi^{(3)\alpha\beta\gamma\delta}_{\mu\nu\rho\sigma}
      \left( - F^{(1)c}_{\gamma\delta} + B^{c}_{\gamma\delta} \right)
\nonumber \\
& + & \frac{\partial {\cal L}_{4}}{\partial F^{(1)a}_{\alpha\beta}}
    + \frac{\partial {\cal L}_{4}}{\partial B^{a}_{\alpha\beta}}
    \equiv 0 .
\end{eqnarray}
They require that the gauge fields $B^{a}_{\mu\nu}$
be incorporated into ${\cal L}_{4}$ in the form of
\begin{equation}
{\cal L}_{4} \equiv
{\cal L}_{5}\left( \varphi, \partial_{\mu}\varphi, \nabla_{\mu\nu}\varphi;
            B_{\rho}, \tilde{F}^{(1)}, F^{(2)},
                      \tilde{C}^{(1)}, \tilde{C}^{(2)};
            \tilde{D}^{(2)}, \tilde{\tilde{G}}^{(2)}, \tilde{G}^{(3)}
     \right) ,
\end{equation}
where
\begin{eqnarray}
\nabla_{\mu\nu}\varphi_{A}
& \definition &
\partial_{\mu\nu}\varphi_{A}
- B^{a}_{\mu\nu}M_{aA}^{\makebox[3mm]{} B}\varphi_{B} , \\
\tilde{F}^{(1)a}_{\alpha\beta}
& \definition &
F^{(1)a}_{\alpha\beta} - B^{a}_{\alpha\beta} , \\
\tilde{C}^{(2)a}_{\alpha\beta\gamma}
& \definition &
C^{(2)a}_{\alpha\beta\gamma}
- f^{a}_{bc}\Gamma^{(2)\mu\nu\rho}_{\alpha\beta\gamma}
                                    B^{b}_{\mu\nu}B^{c}_{\rho} , \\
\tilde{D}^{(2)a}_{\alpha\beta\gamma}
& \definition &
D^{(2)a}_{\alpha\beta\gamma}
+ \frac{1}{2}f^{a}_{bc}\Gamma^{(2)\mu\nu\rho}_{\alpha\beta\gamma}
                                    B^{b}_{\mu\nu}B^{c}_{\rho} , \\
\tilde{\tilde{G}}^{(2)a}_{\alpha\beta\gamma\delta}
& \definition &
\tilde{G}^{(2)a}_{\alpha\beta\gamma\delta}
- f^{a}_{bc}\left(
  \frac{1}{2}\Xi^{(2)\mu\nu\rho\sigma}_{\alpha\beta\gamma\delta}
                                    B^{b}_{\mu\nu}B^{c}_{\rho\sigma}
+ 2\Xi^{(2)\mu\rho\nu\sigma}_{\alpha\beta\gamma\delta}
                                    B^{b}_{\mu\nu}F^{(2)c}_{\rho\sigma}
\right) \nonumber \\
& & \makebox[1cm]{}
+ f^{e}_{bc}f^{a}_{ed}
  \Xi^{(2)\xi\eta\zeta\sigma}_{\alpha\beta\gamma\delta}
  \Gamma^{(1)\mu\nu\rho}_{\xi\eta\zeta}
  B^{b}_{\mu\nu}B^{c}_{\rho}B^{d}_{\sigma} , \\
\tilde{G}^{(3)a}_{\alpha\beta\gamma\delta}
& \definition &
G^{(3)a}_{\alpha\beta\gamma\delta}
+ f^{a}_{bc}\Xi^{(3)\mu\nu\rho\sigma}_{\alpha\beta\gamma\delta}
                                    B^{b}_{\mu\nu}F^{(1)c}_{\rho\sigma} .
\end{eqnarray}

The Lagrangian ${\cal L}_{5}$ gives the identities (28) the following form:
\begin{eqnarray}
\lefteqn{
\frac{\partial {\cal L}_{5}}{\partial\partial_{\alpha}\varphi_{A}}
M_{aA}^{\makebox[3mm]{} B}\varphi_{B}
+ 2\frac{\partial {\cal L}_{5}}{\partial\nabla_{\alpha\mu}\varphi_{A}}
\left( M_{aA}^{\makebox[3mm]{} B}\partial_{\mu}\varphi_{B}
     - \frac{1}{2}f^{b}_{ac}
       M_{bA}^{\makebox[3mm]{} B}B^{c}_{\mu}\varphi_{B} \right)
}\nonumber \\
& + & \frac{\partial {\cal L}_{5}}{\partial F^{(2)b}_{\mu\nu}}
      f^{b}_{ac}\Phi^{(2)\alpha\beta}_{\mu\nu}B^{c}_{\beta}
\nonumber \\
& + & \frac{\partial {\cal L}_{5}}{\partial \tilde{C}^{(1)b}_{\mu\nu\rho}}
      f^{b}_{ac}\Gamma^{(1)\alpha\beta\gamma}_{\mu\nu\rho}
      \tilde{F}^{(1)c}_{\alpha\beta}
\nonumber \\
& + & 2\frac{\partial {\cal L}_{5}}{\partial \tilde{C}^{(2)b}_{\mu\nu\rho}}
      \left\{ f^{b}_{ac}\Gamma^{(2)\alpha\beta\gamma}_{\mu\nu\rho}
              \left( \tilde{F}^{(1)c}_{\beta\gamma} + F^{(2)c}_{\beta\gamma}
              \right)
            - \frac{1}{2}f^{e}_{ac}f^{b}_{ed}
              \Gamma^{(2)\alpha\beta\gamma}_{\mu\nu\rho}
              B^{c}_{\beta}B^{d}_{\gamma} \right\}
\nonumber \\
& + & \frac{\partial {\cal L}_{5}}{\partial \tilde{D}^{(2)b}_{\mu\nu\rho}}
      \left\{ f^{b}_{ac}\Delta^{(2)\beta\gamma\alpha}_{\mu\nu\rho}
              \left( \tilde{F}^{(1)c}_{\beta\gamma} + F^{(2)c}_{\beta\gamma}
              \right)
            + \frac{1}{2}f^{e}_{ac}f^{b}_{ed}
              \Gamma^{(2)\alpha\beta\gamma}_{\mu\nu\rho}
              B^{c}_{\beta}B^{d}_{\gamma} \right\}
\nonumber \\
& + & \frac{\partial {\cal L}_{5}}
               {\partial \tilde{\tilde{G}}^{(2)b}_{\mu\nu\rho\sigma}}
     \left\{ f^{b}_{ac}\Xi^{(2)\alpha\beta\gamma\delta}_{\mu\nu\rho\sigma}
             \left( - \tilde{C}^{(1)c}_{\beta\gamma\delta}
                    + \frac{8}{3}\tilde{C}^{(2)c}_{\beta\gamma\delta}
                    + \frac{8}{3}\tilde{D}^{(2)c}_{\gamma\delta\beta}
             \right) \right.
\nonumber \\
& & \makebox[15mm]{}
           + f^{e}_{da}f^{b}_{ec}
             \Xi^{(2)\xi\eta\zeta\beta}_{\mu\nu\rho\sigma}
             \Gamma^{(1)\alpha\gamma\delta}_{\xi\eta\zeta}
             B^{c}_{\beta}\tilde{F}^{(1)d}_{\gamma\delta}
           - 2f^{e}_{ac}f^{b}_{ed}
             \Xi^{(2)\alpha\gamma\beta\delta}_{\mu\nu\rho\sigma}
             B^{c}_{\beta}F^{(2)d}_{\gamma\delta}
\nonumber \\
& & \makebox[15mm]{}
    + \left. f^{f}_{ac}f^{g}_{fd}f^{b}_{ge}
             \Xi^{(2)\xi\eta\zeta\delta}_{\mu\nu\rho\sigma}
             \Gamma^{(1)\alpha\beta\gamma}_{\xi\eta\zeta}
             B^{c}_{\beta}B^{d}_{\gamma}B^{e}_{\delta} \right\}
\nonumber \\
& + & \frac{\partial {\cal L}_{5}}
               {\partial \tilde{G}^{(3)b}_{\mu\nu\rho\sigma}}
      \left\{ - \frac{8}{3}f^{b}_{ac}
                \Xi^{(3)\alpha\beta\gamma\delta}_{\mu\nu\rho\sigma}
                \left( \tilde{C}^{(2)c}_{\beta\gamma\delta}
                     - \tilde{D}^{(2)c}_{\gamma\delta\beta}
                \right)
              + f^{e}_{ac}f^{b}_{ed}
                \Xi^{(3)\alpha\beta\gamma\delta}_{\mu\nu\rho\sigma}
                B^{c}_{\beta}\tilde{F}^{(1)d}_{\gamma\delta} \right\}
\nonumber \\
& + & \frac{\partial {\cal L}_{5}}{\partial B^{a}_{\alpha}}
\equiv 0 .
\end{eqnarray}
{}From them we find that the gauge fields $B^{a}_{\mu}$ should be contained
as follows:
\begin{equation}
{\cal L}_{5} \equiv
{\cal L}_{6}\left(
            \varphi, \nabla_{\mu}\varphi, \tilde{\nabla}_{\mu\nu}\varphi;
            \tilde{F}^{(1)}, \tilde{F}^{(2)},
                   \tilde{\tilde{C}}^{(1)}, \tilde{\tilde{C}}^{(2)};
            \tilde{\tilde{D}}^{(2)},
                   \tilde{\tilde{\tilde{G}}}^{(2)}, \tilde{\tilde{G}}^{(3)}
     \right) ,
\end{equation}
where
\begin{eqnarray}
\nabla_{\mu}\varphi_{A}
& \definition &
\partial_{\mu}\varphi_{A}
- B^{a}_{\mu}M_{aA}^{\makebox[3mm]{} B}\varphi_{B} , \\
\tilde{\nabla}_{\mu\nu}\varphi_{A}
& \definition &
\nabla_{\mu\nu}\varphi_{A}
- \left( B^{a}_{\mu}M_{aA}^{\makebox[3mm]{} B}\partial_{\nu}\varphi_{B}
       + B^{a}_{\nu}M_{aA}^{\makebox[3mm]{} B}\partial_{\mu}\varphi_{B}
  \right)
\nonumber \\
& & \makebox[13mm]{}
+ \frac{1}{2}B^{a}_{\mu}B^{b}_{\nu}
  \{ M_{a}, M_{b} \} _{A}^{\makebox[1mm]{} B}\varphi_{B}
, \\
\tilde{F}^{(2)a}_{\mu\nu}
& \definition &
F^{(2)a}_{\mu\nu} - \frac{1}{2}f^{a}_{bc}B^{b}_{\mu}B^{c}_{\nu} , \\
\tilde{\tilde{C}}^{(1)a}_{\mu\nu\rho}
& \definition &
\tilde{C}^{(1)a}_{\mu\nu\rho}
- f^{a}_{bc}\Gamma^{(1)\alpha\beta\gamma}_{\mu\nu\rho}
            B^{b}_{\alpha}\tilde{F}^{(1)c}_{\beta\gamma} , \\
\tilde{\tilde{C}}^{(2)a}_{\mu\nu\rho}
& \definition &
\tilde{C}^{(2)a}_{\mu\nu\rho}
- 2f^{a}_{bc}\Gamma^{(2)\alpha\beta\gamma}_{\mu\nu\rho}
             B^{b}_{\alpha}
             \left( \tilde{F}^{(1)c}_{\beta\gamma}
                  + F^{(2)c}_{\beta\gamma} \right)
+ \frac{2}{3}f^{e}_{bc}f^{a}_{ed}
             \Gamma^{(2)\beta\gamma\alpha}_{\mu\nu\rho}
             B^{b}_{\alpha}B^{c}_{\beta}B^{d}_{\gamma} , \\
\tilde{\tilde{D}}^{(2)a}_{\mu\nu\rho}
& \definition &
\tilde{D}^{(2)a}_{\mu\nu\rho}
- f^{a}_{bc}\Delta^{(2)\beta\gamma\alpha}_{\mu\nu\rho}
             B^{b}_{\alpha}
             \left( \tilde{F}^{(1)c}_{\beta\gamma}
                  + F^{(2)c}_{\beta\gamma} \right)
- \frac{1}{3}f^{e}_{bc}f^{a}_{ed}
             \Gamma^{(2)\beta\gamma\alpha}_{\mu\nu\rho}
             B^{b}_{\alpha}B^{c}_{\beta}B^{d}_{\gamma} , \\
\tilde{\tilde{\tilde{G}}}^{(2)a}_{\mu\nu\rho\sigma}
& \definition &
\tilde{\tilde{G}}^{(2)a}_{\mu\nu\rho\sigma}
- f^{a}_{bc}\Xi^{(2)\alpha\beta\gamma\delta}_{\mu\nu\rho\sigma}
  B^{b}_{\alpha}
  \left( - \tilde{C}^{(1)c}_{\beta\gamma\delta}
         + \frac{8}{3}\tilde{C}^{(2)c}_{\beta\gamma\delta}
         + \frac{8}{3}\tilde{D}^{(2)c}_{\gamma\delta\beta} \right)
\nonumber \\
& & \makebox[1cm]{}
- 2f^{e}_{bd}f^{a}_{ec}
  \Xi^{(2)\alpha\gamma\beta\delta}_{\mu\nu\rho\sigma}
  B^{b}_{\alpha}B^{c}_{\beta}F^{(2)d}_{\gamma\delta}
\nonumber \\
& & \makebox[1cm]{}
- \frac{1}{2}f^{f}_{bd}f^{g}_{fc}f^{a}_{ge}
  \Xi^{(2)\alpha\beta\gamma\delta}_{\mu\nu\rho\sigma}
  B^{b}_{\alpha}B^{c}_{\beta}B^{d}_{\gamma}B^{e}_{\delta} , \\
\tilde{\tilde{G}}^{(3)a}_{\mu\nu\rho\sigma}
& \definition &
\tilde{G}^{(3)a}_{\mu\nu\rho\sigma}
+ \frac{8}{3}f^{a}_{bc}
  \Xi^{(3)\alpha\beta\gamma\delta}_{\mu\nu\rho\sigma}
  B^{b}_{\alpha}
  \left( \tilde{C}^{(2)c}_{\beta\gamma\delta}
       - \tilde{D}^{(2)c}_{\gamma\delta\beta} \right)
\nonumber \\
& & \makebox[1cm]{}
+ f^{e}_{bd}f^{a}_{ec}
  \Xi^{(3)\alpha\beta\gamma\delta}_{\mu\nu\rho\sigma}
  B^{b}_{\alpha}B^{c}_{\beta}\tilde{F}^{(1)d}_{\gamma\delta} .
\end{eqnarray}

The Lagrangian ${\cal L}_{6}$ should satisfy the last
identities (27) rewritten as
\begin{eqnarray}
\lefteqn{ \frac{\partial {\cal L}_{6}}{\partial\varphi_{A}}
  M_{aA}^{\makebox[3mm]{} B}\varphi_{B}
+ \frac{\partial {\cal L}_{6}}{\partial\nabla_{\mu}\varphi_{A}}
  M_{aA}^{\makebox[3mm]{} B}\nabla_{\mu}\varphi_{B}
+ \frac{\partial {\cal L}_{6}}{\partial\tilde{\nabla}_{\mu\nu}\varphi_{A}}
  M_{aA}^{\makebox[3mm]{} B}\tilde{\nabla}_{\mu\nu}\varphi_{B}
}\nonumber \\
& + & \frac{\partial {\cal L}_{6}}{\tilde{F}^{(1)b}_{\mu\nu}}
      f^{b}_{ac}\tilde{F}^{(1)c}_{\mu\nu}
    + \frac{\partial {\cal L}_{6}}{\tilde{F}^{(2)b}_{\mu\nu}}
      f^{b}_{ac}\tilde{F}^{(2)c}_{\mu\nu}
    + \frac{\partial {\cal L}_{6}}{\tilde{\tilde{C}}^{(1)b}_{\mu\nu\rho}}
      f^{b}_{ac}\tilde{\tilde{C}}^{(1)c}_{\mu\nu\rho}
    + \frac{\partial {\cal L}_{6}}{\tilde{\tilde{C}}^{(2)b}_{\mu\nu\rho}}
      f^{b}_{ac}\tilde{\tilde{C}}^{(2)c}_{\mu\nu\rho}
\nonumber \\
& + & \frac{\partial {\cal L}_{6}}{\tilde{\tilde{D}}^{(2)b}_{\mu\nu\rho}}
      f^{b}_{ac}\tilde{\tilde{D}}^{(2)c}_{\mu\nu\rho}
    + \frac{\partial {\cal L}_{6}}
               {\tilde{\tilde{\tilde{G}}}^{(2)b}_{\mu\nu\rho\sigma}}
      f^{b}_{ac}\tilde{\tilde{\tilde{G}}}^{(2)c}_{\mu\nu\rho\sigma}
    + \frac{\partial {\cal L}_{6}}
               {\tilde{\tilde{G}}^{(3)b}_{\mu\nu\rho\sigma}}
      f^{b}_{ac}\tilde{\tilde{G}}^{(3)c}_{\mu\nu\rho\sigma}
\equiv 0 .
\end{eqnarray}
They show the Lagrangian ${\cal L}_{6}$ should be an
arbitrary $G$-invariant function of the arguments.
This is our final result.

In the above we have determined the covariant derivatives and gauge-field
strengths step by step.
For completeness we express them by using the original gauge fields and
the irreducible components of their derivatives:
\begin{eqnarray}
\nabla_{\mu}\varphi_{A}
& = &
\partial_{\mu}\varphi_{A}
- B^{a}_{\mu}M_{aA}^{\makebox[3mm]{} B}\varphi_{B} , \\
\tilde{\nabla}_{\mu\nu}\varphi_{A}
& = &
\partial_{\mu\nu}\varphi_{A}
- B^{a}_{\mu\nu}M_{aA}^{\makebox[3mm]{} B}
- \left( B^{a}_{\mu}M_{aA}^{\makebox[3mm]{} B}\partial_{\nu}\varphi_{B}
       + B^{a}_{\nu}M_{aA}^{\makebox[3mm]{} B}\partial_{\mu}\varphi_{B}
  \right)
\nonumber \\
& & \makebox[4cm]{}
+ \frac{1}{2}B^{a}_{\mu}B^{b}_{\nu}
  \{ M_{a}, M_{b} \} _{A}^{\makebox[1mm]{} B}\varphi_{B}
, \\
\tilde{F}^{(1)a}_{\mu\nu}
& = &
F^{(1)a}_{\mu\nu} - B^{a}_{\mu\nu} , \\
\tilde{F}^{(2)a}_{\mu\nu}
& = &
F^{(2)a}_{\mu\nu} - \frac{1}{2}f^{a}_{bc}B^{b}_{\mu}B^{c}_{\nu} , \\
\tilde{\tilde{C}}^{(1)a}_{\mu\nu\rho}
& = &
C^{(1)a}_{\mu\nu\rho} - D^{(1)a}_{\mu\nu\rho}
- f^{a}_{bc}\Gamma^{(1)\alpha\beta\gamma}_{\mu\nu\rho}
            B^{b}_{\alpha}F^{(1)c}_{\beta\gamma}
+ f^{a}_{bc}\Gamma^{(1)\alpha\beta\gamma}_{\mu\nu\rho}
            B^{b}_{\alpha}B^{c}_{\beta\gamma} , \\
\tilde{\tilde{C}}^{(2)a}_{\mu\nu\rho}
& = &
C^{(2)a}_{\mu\nu\rho}
- 2f^{a}_{bc}\Gamma^{(2)\alpha\beta\gamma}_{\mu\nu\rho}
             B^{b}_{\alpha}
             \left( F^{(1)c}_{\beta\gamma}
                  + F^{(2)c}_{\beta\gamma} \right)
+ \frac{2}{3}f^{a}_{be}f^{e}_{cd}
             \Gamma^{(2)\alpha\beta\gamma}_{\mu\nu\rho}
             B^{b}_{\alpha}B^{c}_{\beta}B^{d}_{\gamma} , \\
\tilde{\tilde{D}}^{(2)a}_{\mu\nu\rho}
& = &
D^{(2)a}_{\mu\nu\rho}
- f^{a}_{bc}\Gamma^{(2)\alpha\beta\gamma}_{\mu\nu\rho}
             B^{b}_{\alpha}
             \left( F^{(1)c}_{\beta\gamma}
                  - F^{(2)c}_{\beta\gamma} \right)
+ 2f^{a}_{bc}\Gamma^{(2)\alpha\beta\gamma}_{\mu\nu\rho}
             B^{b}_{\alpha}B^{c}_{\beta\gamma}
\nonumber \\
& & \makebox[64mm]{}
- \frac{1}{3}f^{a}_{be}f^{e}_{cd}
             \Gamma^{(2)\alpha\beta\gamma}_{\mu\nu\rho}
             B^{b}_{\alpha}B^{c}_{\beta}B^{d}_{\gamma} , \\
\tilde{\tilde{\tilde{G}}}^{(2)a}_{\mu\nu\rho\sigma}
& = &
G^{(2)a}_{\mu\nu\rho\sigma}
- 2f^{a}_{bc}\Xi^{(2)\alpha\gamma\beta\delta}_{\mu\nu\rho\sigma}
  B^{b}_{\alpha\beta}F^{(2)c}_{\gamma\delta}
+ 2f^{a}_{be}f^{e}_{cd}\Xi^{(2)\alpha\gamma\beta\delta}_{\mu\nu\rho\sigma}
  B^{b}_{\alpha}B^{c}_{\beta}F^{(2)d}_{\gamma\delta}
\nonumber \\
& & \makebox[1cm]{}
+ f^{a}_{bc}\Xi^{(2)\alpha\beta\gamma\delta}_{\mu\nu\rho\sigma}
  B^{b}_{\alpha}
  \left( C^{(1)c}_{\beta\gamma\delta}
         - \frac{8}{3}C^{(2)c}_{\beta\gamma\delta}
         - 2D^{(1)c}_{\beta\gamma\delta}
         + \frac{16}{3}D^{(2)c}_{\beta\gamma\delta} \right)
\nonumber \\
& & \makebox[1cm]{}
- \frac{1}{2}f^{a}_{bc}\Xi^{(2)\alpha\beta\gamma\delta}_{\mu\nu\rho\sigma}
  B^{b}_{\alpha\beta}B^{c}_{\gamma\delta}
+ f^{a}_{be}f^{e}_{cd}
  \left( \Xi^{(2)\alpha\beta\gamma\delta}_{\mu\nu\rho\sigma}
       - 2\Xi^{(2)\alpha\gamma\beta\delta}_{\mu\nu\rho\sigma} \right)
  B^{b}_{\alpha}B^{c}_{\beta}B^{d}_{\gamma\delta}
\nonumber \\
& & \makebox[1cm]{}
- \frac{1}{2}f^{a}_{bg}f^{g}_{cf}f^{f}_{de}
  \Xi^{(2)\alpha\gamma\beta\delta}_{\mu\nu\rho\sigma}
  B^{b}_{\alpha}B^{c}_{\beta}B^{d}_{\gamma}B^{e}_{\delta} , \\
\tilde{\tilde{G}}^{(3)a}_{\mu\nu\rho\sigma}
& = &
G^{(3)a}_{\mu\nu\rho\sigma}
+ f^{a}_{bc}\Xi^{(3)\alpha\beta\gamma\delta}_{\mu\nu\rho\sigma}
  B^{b}_{\alpha\beta}F^{(1)c}_{\gamma\delta}
- f^{a}_{be}f^{e}_{cd}
  \Xi^{(3)\alpha\beta\gamma\delta}_{\mu\nu\rho\sigma}
  B^{b}_{\alpha}B^{c}_{\beta}F^{(1)d}_{\gamma\delta}
\nonumber \\
& & \makebox[1cm]{}
+ \frac{8}{3}f^{a}_{bc}
  \Xi^{(3)\alpha\beta\gamma\delta}_{\mu\nu\rho\sigma}
  B^{b}_{\alpha}
  \left( C^{(2)c}_{\beta\gamma\delta}
       + 2D^{(2)c}_{\beta\gamma\delta} \right)
\nonumber \\
& & \makebox[1cm]{}
+ f^{a}_{be}f^{e}_{cd}
  \Xi^{(3)\alpha\beta\gamma\delta}_{\mu\nu\rho\sigma}
  B^{b}_{\alpha}B^{c}_{\beta}B^{d}_{\gamma\delta} .
\end{eqnarray}

\section{Summary and Discussion}

In the present paper we have generalized the usual prescription
for constructing gauge-invariant Lagrangian
to the case of including second derivatives of fields as well as first
derivatives.
By solving a series of identities which follow from generalized
Noether's theorems,
we have found the covariant derivatives and the gauge-field strengths.

Many problems remain to be solved: \\
---physical implications of the tensor gauge fiels; \\
---geometrical meanings of gauging second derivatives; \\
---extension to space-time symmetries; \\
---higher-derivative gravity in the light of this new type of gauge theories.

\section*{Acknowledgments}

The author would like to thank Takanori Fujiwara and Minoru Hirayama
for discussions.

\end{document}